# Analysis and Classification of Critical Success Factors for Business Conversation Manager (BCM) System Implementation in Service-Oriented Enterprises

**Safiollah Heidari[1][*], Mohammad Jafar Tarokh[2]**

[1]IT Group, Industrial Engineering Department, K. N. Toosi University of Technology, Tehran, Iran, safi.heidari@sina.kntu.ac.ir

[2]IT Group, Industrial Engineering Department, K. N. Toosi University of Technology, Tehran, Iran, mjtarokh@kntu.ac.ir

## Abstract

Business conversation manager is a system which makes organization's communications possible on computer system platforms and also within defined processes and relationships. This research reviews the literature of BCM systems and introduces and classifies the most important factors for successful implementation by investigating various references. The aim is to identify critical success factors in order to avoid high risk probabilities and facilitate the implementation and finally getting better outcomes. Next, factors are extracted from experts' points of view and they depicted as a model to show critical factors.

## 1. Introduction

Business conversation manager (BCM) is a system which makes organization's communications possible on computer system platforms and also within defined processes and relationships. This system facilitates reaching organizational goals and assists managers to make appropriate decision for various situations. Business conversation expression refers to a conceptual container for interaction between organizations to achieve business goals [1] which are shared among them. On the other hand, an agenda or a process itself contains several tasks which there are many relationships between them [2]. These tasks will be gathered and work in the form of different services and they will be coordinated by orchestration. Also, each conversation includes several participants that may be a real people or a role. In addition, a number of documents are manufactured or used in conversations as inputs or outputs [1]. Finally, a conversation, with all mentioned characteristics, uses the agenda for achieving organization's goals.

---

[1] *Corresponding author:*
Safiollah Heidari (safi.heidari@sina.kntu.ac.ir)

Implementing such a system is not so simple. Before implementing such a system in action in the organization, a series of preparations should be provided in all areas that it is involved. On the other hand, gradually over time, organizations can no longer ignore the benefits of having such a system because extensive relationships in organizations, the need for integrating relationships between outside and inside the organization, relationships between business partners and even competitors, the need for fast and consistent relationships and many other factors associated with an increased scope of business make using this system inevitable to maintain the organization's competitive advantages.

Most information systems that are used in organizations are not flexible enough. They are just doing a series of predefined actions and have not performance beyond the framework which is defined for them. While a considerable note in organizational relationships is informal conversations and interactions between different processes and roles [1]. These informal channels may include e-mail, phone, etc. They are important sources of decisions. One advantage of business conversation manager (BCM) system is that beside formal communications that are performed in the letters, instructions, notes, etc., it supports informal conversations and communications too.

In this research, we considered BCM beyond the current systems. As in addition to re-engineering that must be applied to the processes, X-engineering concepts are considered too. It means that BCM, in addition to covering the ongoing processes within the organization, will contains anyone who is in contact with the organization, including customers, suppliers and partners [3]. Even competitors are covered by this system. As in today's world where all relationships are interdependent, interacting appropriately with competitors in order to achieve the organization's strategic objectives cannot be ignored.

The important point here is that implementing such a system requires some preparation on behalf of the organization. For example, the organization that wants to implement this system should at least done business process re-engineering (BPR) once before. This will facilitate and expedite the identification of strength and weaknesses of the system for implementing. Thus, examining the factors that lead to successful implementation of BCM system and reduce risks in the organization is critical. On the other hand, failure to identify the success factors and a correct classification in this context is one of the problems that can waste a lot of time and money and finally does not include any results.

Generally, critical success factors are used to identify information required for listing and describing the essential elements for systems and application success in order to help to define them and focus on the efforts and responsibilities of management [15]. "Success factors are limited number of areas that will be satisfactory if they enhance the competitive ability in the organization. These are some areas that will improve business and if you have not done enough research on them, the organization's effort to achieve success may be undesirable." Griffin and Kenneth said [16]]. "Success factors are activity contexts that need special attention on behalf of the management. The current state of performance of these activities should be continuously evaluated and that information should always be available." Rockart said [17].

In the following, factors affecting the successful implementation of BCM are classified and then with a survey conducted by the experts, the most important and influential factors for each group are determined.

## 2. Critical success factors for BCM implementation

After a series of surveys, critical success factors for BCM implementation were classified into four groups that each of them contains a number of success factors that affect implementation of the system. The four groups are:

*A. Issues related to organizational structure and plans*

Organizational structure and its problems is one of the most important issues that must be addressed. As mentioned, one of the essential points that should be considered is that organization's processes must be re-engineered at least once. Plans that are predicted in the organization's strategies and instructions are also very important.

*B. Issues related to organizational culture and leadership*

Considering the issues related to organizational culture and leadership are other important issues that are critical for reducing risks in the project.

*C. Issues related to communication and system security*

Because of the breadth of processes and roles that are covered by this system, security issues such as confidentiality, integrity and availability of data and information and appropriate authentication are of particular importance. Besides, good communication areas should be considered too.

*D. Issues related to the techniques used in the system*

Most of these issues pay attention to the methods that have been used to establish good communication and type of system design with respect to all aspects of organizational structure and relationships.

Most considerable success factors in each group are listed as follows:

Table1. Overview of key success factors for BCM system implementation

| Organizational structure and plans factors | Organizational culture and leadership factors |
|---|---|
| Change the organization's structure from task-oriented to process-oriented [3] | Changing the organization's culture from the traditional to the technology-based culture[3] |
| Improve communication processes with suppliers [3] | Staff training [5] |
| Hierarchical levels of reference documents[4] | Motivating employees to transfer their activities to the system |
| Transparent processes for referral documents [7] | Senior management commitment to implement BCM system [5] |
| Standardized documents of communication with suppliers [1] | Senior management support of the project costs [5] |
| Clarify the processes' responsibilities [8] | Business commitment to BCM systems process and procedures [10] |
| True understanding of processes [8] | Managerial incentives to enter into this realm [6] |
| Implement BPR before designing the system [1] | Encourage all areas involved in the organization to use the system(employees, suppliers, customers, etc.) |
| Plans to cope with unexpected changes before, during and after implementation | Justify the areas involved in BCM system and emphasize of its competitive advantages |
| Proper planning for BCM implementation [10] | |
| Proper and efficient management of resources involved in the project [10] | |
| Plenty of skills among the members involved in the project | |
| **Communication and system security** | **The techniques used in the system factors** |
| Determining the level of authorizing and access [13] | Suitable hardware, software and networking platforms (Infrastructures) [4] |
| Creating appropriate controls to maintain data integrity [6] | Choose appropriate method for implementation (Big-bang or Step-by-step) [5] |
| Communication between system designers and staff | Using information technology requirements (e-mail, storage devices, etc.) [1] |
| Proper definition for changes scope [4] | Object-oriented design of software package [5][11] |
| Use of external consultants | Proportion of software design with organization's processes [1] |
| True understanding of customers' relationship requirements [3] | Support for internal conversations and talks [1][2] |
| True understanding of requirements for relationship with suppliers [3] | Flexibility to adapt the system to the processes [9] |
| True understanding of requirements for relationship with partners [3] | Providing suitable manuals and catalogs for the system |
| Using appropriate communication channels [11][12] | Designing plans for system maintenance (e.g. backup database) [1][2] |

| | Using the right tools and tests to evaluate the effects of the designed system, before running [5] |
|---|---|

## 3. Analysis of critical success factors for BCM system implementation

After collecting the key success factors for BCM implementation project, a questionnaire was designed using Likert [14] method and it was sent to 263 professors, PhD students and scientists from reputable companies in person or via email from inside and outside of Iran. These people were involved in implementation and enforcement of BCM systems and have enough experiences in this field.

For this purpose, experts rated each factor independent according to the importance of them in successful BCM system implementation. Likert scales include:

1) Very little effect
2) Weak effect
3) Average effect
4) Significant effect
5) Extremely significant effect

The questionnaire was designed in two levels. In the first stage, major groups were rated in the degree of importance. Then in second stage, subsidiary factors were rated. After scrutinizing responses, the answers which were invalid or incomplete, were omitted and finally 226 questionnaires were specified as valid answers. The most important success factors for implementation of BCM system were extracted and prioritized by distinguishing experts' answers. To do this, answers at each level were calculated and the averages were computed for each level. Because the aim was to identify the most important and influential critical success factors for BCM system implementation, subsidiary factors with 4 score and 5 score were taken into consideration. Figure 1 shows the research workflow diagram and figure 2 shows the priorities obtained from answers. Because of all these factors cannot be considered, the most important factors that have the greatest impact on successful implementation are listed.

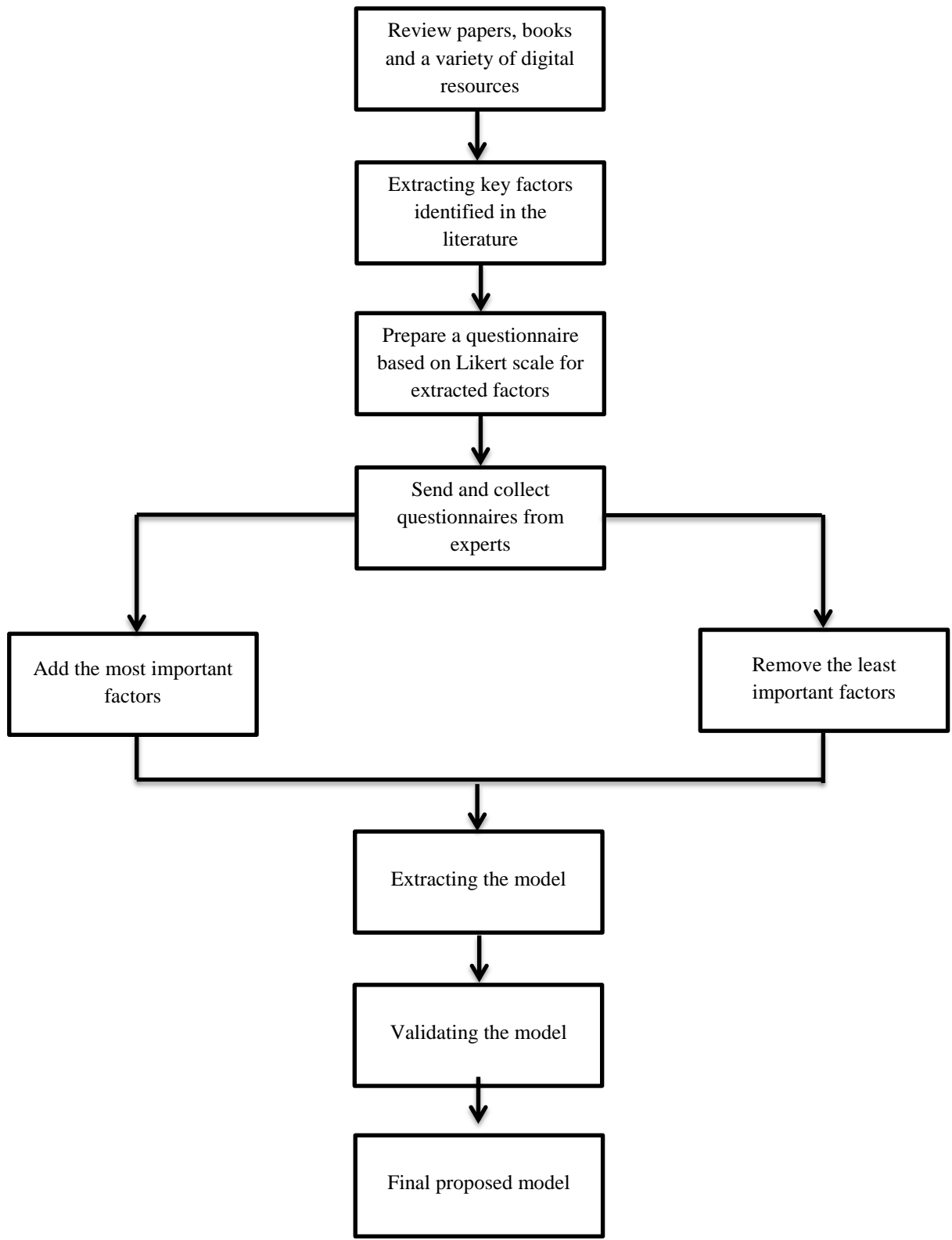

Figure1. The research process

The factors are divided into sub-groups in each category, as is shown in figure 4, in order to use and understanding of each factor in real environments be more tangible. In the following tables, the frequency answers of respondents are calculated using SPSS v.11.5 software and have been displayed. To determine the reliability of the questionnaire, Cronbach alpha coefficient was used. In this study the alpha was 91% that is a desired value. These tables have been divided into four groups, which are located in the first section of the questionnaire.

Table2. Organizational structure and plans factors

| Factors | Avg. | Standard deviation | Variation |
|---|---|---|---|
| Change the organization's structure from task-oriented to process-oriented | 4.59 | 0.498 | 0.108 |
| Improve communication processes with suppliers | 3.46 | 0.802 | 0.231 |
| Hierarchical levels of reference documents | 3.75 | 0.672 | 0.179 |
| Transparent processes for referral documents | 4.65 | 0.482 | 0.103 |
| Standardized documents of communication with suppliers | 4.21 | 0.420 | 0.099 |
| Clarify the processes' responsibilities | 4.71 | 0.456 | 0.096 |
| True understanding of processes | 4.75 | 0.439 | 0.092 |
| Implement BPR before designing the system | 3.93 | 0.759 | 0.193 |
| Plans to cope with unexpected changes before, during and after implementation | 3.68 | 0.737 | 0.200 |
| Proper planning for BCM implementation | 3.87 | 0.707 | 0.182 |
| Proper and efficient management of resources involved in the project | 3.43 | 0.564 | 0.164 |
| Plenty of skills among the members involved in the project | 4.34 | 0.482 | 0.111 |

Table3. Organizational culture and leadership factors

| Factors | Avg. | Standard deviation | Variation |
|---|---|---|---|
| Changing the organization's culture from the traditional to the technology-based culture | 4.56 | 0.504 | 0.110 |
| Staff training | 3.34 | 0.482 | 0.144 |
| Motivating employees to transfer their activities to the system | 3.71 | 0.924 | 0.249 |
| Senior management commitment to implement BCM system | 3.87 | 0.832 | 0.214 |
| Senior management support of the project costs | 3.87 | 0.832 | 0.214 |
| Business commitment to BCM systems process and procedures | 4.28 | 0.456 | 0.106 |
| Managerial incentives to enter into this realm | 4.88 | 0.296 | 0.060 |
| Encourage all areas involved in the organization to use the system (employees, suppliers, customers, etc.) | 4.22 | 0.707 | 0.167 |
| Justify the areas involved in BCM system and emphasize of its competitive advantages | 4.62 | 0.491 | 0.106 |

Table4. Communication and system security factors

| Factors | Avg. | Standard deviation | Variation |
|---|---|---|---|
| Determining the level of authorizing and access | 4.43 | 0.504 | 0.113 |
| Creating appropriate controls to maintain data integrity | 3.84 | 0.677 | 0.176 |
| Communication between system designers and staff | 3.81 | 0.780 | 0.204 |
| Proper definition for changes scope | 4.81 | 0.396 | 0.082 |
| Use of external consultants | 3.96 | 0.736 | 0.185 |
| True understanding of customers' relationship requirements | 4.59 | 0.498 | 0.108 |
| True understanding of requirements for relationship with suppliers | 4.28 | 0.456 | 0.106 |
| True understanding of requirements for relationship with partners | 4.15 | 0.723 | 0.174 |
| Using appropriate communication channels | 4.40 | 0.498 | 0.113 |

Table5. The techniques used in the system

| Factors | Avg. | Standard deviation | Variation |
|---|---|---|---|
| Suitable hardware, software and networking platforms (Infrastructures) | 4.25 | 0.567 | 0.133 |
| Choose appropriate method for implementation (Big-bang or Step-by-step) | 4.31 | 0.692 | 0.160 |
| Using information technology requirements (e-mail, storage devices, etc.) | 4.87 | 0.336 | 0.068 |
| Object-oriented design of software package | 3.43 | 0.800 | 0.233 |
| Proportion of software design with organization's processes | 3.96 | 0.897 | 0.226 |
| Support for internal conversations and talks | 3.87 | 0.941 | 0.243 |
| Flexibility to adapt the system to the processes | 4.65 | 0.482 | 0.103 |
| Providing suitable manuals and catalogs for the system | 3.87 | 0.832 | 0.214 |
| Designing plans for system maintenance (e.g. backup database) | 3.43 | 0.564 | 0.164 |
| Using the right tools and tests to evaluate the effects of the designed system, before running | 3.34 | 0.482 | 0.144 |

## 4. Conclusion

Conversation-oriented systems are new and innovative technologies that will gradually find their place in organizational processes and organizations should inevitably use these systems to maintain their competitive advantages. Therefore, identifying, analyzing and understanding the most important factors for implementing those systems can reduce risks and also it can increase the likelihood of successful implementation. One of the most important conversation-oriented systems is business conversation manager (BCM) which covers most parts of the organization. It also acts as a facilitator of human agents' internal and external working communication in the organization, while enables define, modify and improve participatory processes in which participants work interactively.

This research reviews the literature of critical success factors for BCM system implementation to identify and classify the most important factors. Among the key benefits is that managers can identify success factors and have a greater mastery. Classified factors made clear what factors should be considered in what part of the organization. According to the extracted results, risks rated in BCM system implementation project can be reduced significantly.

## Acknowledgment

The authors would like to thank Dr. Hamid Motahari-Nezhad from Hewlett-Packard Lab. for providing us many useful guidance and data.

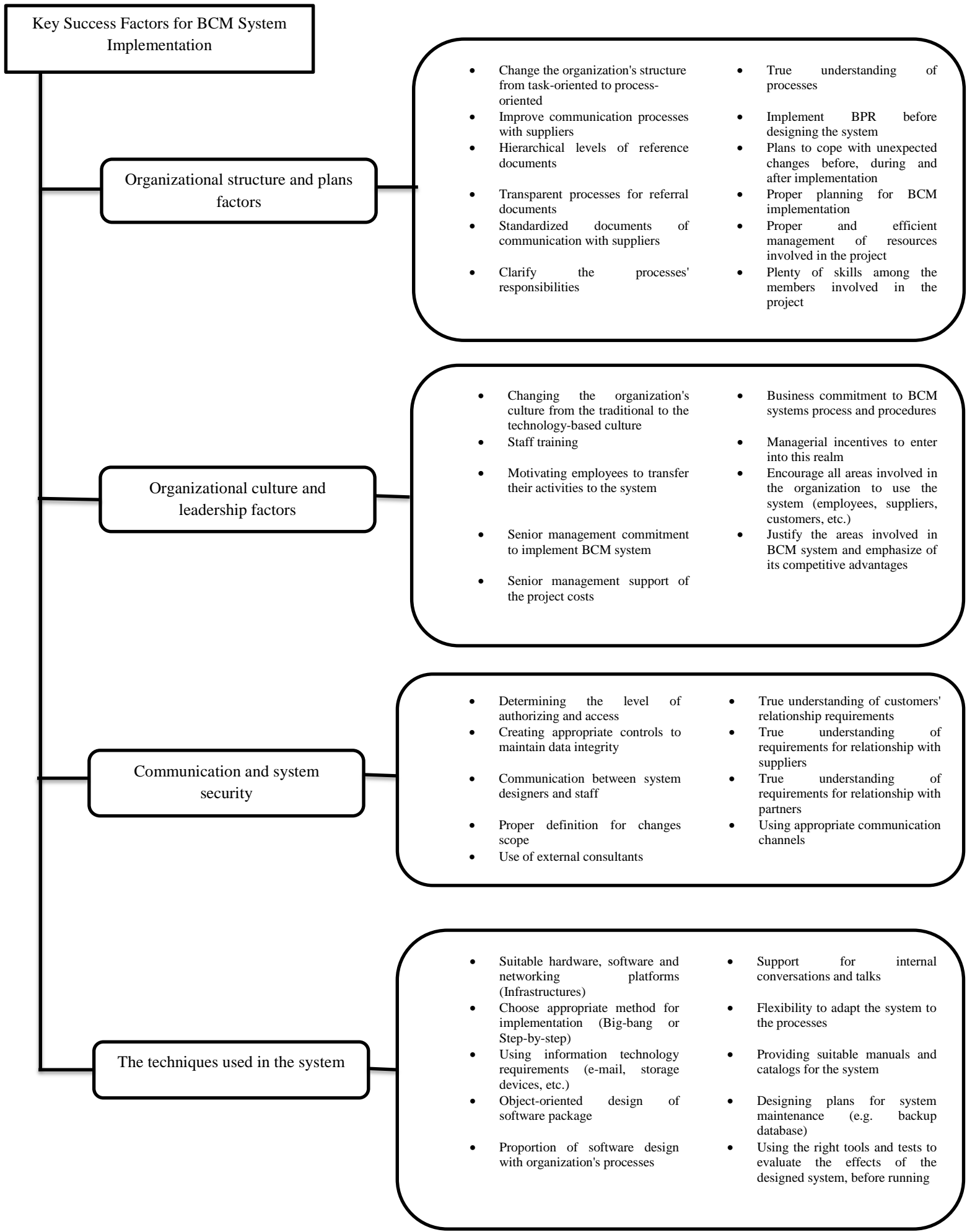

Figure2. Tree diagram of key success factors for BCM system implementation

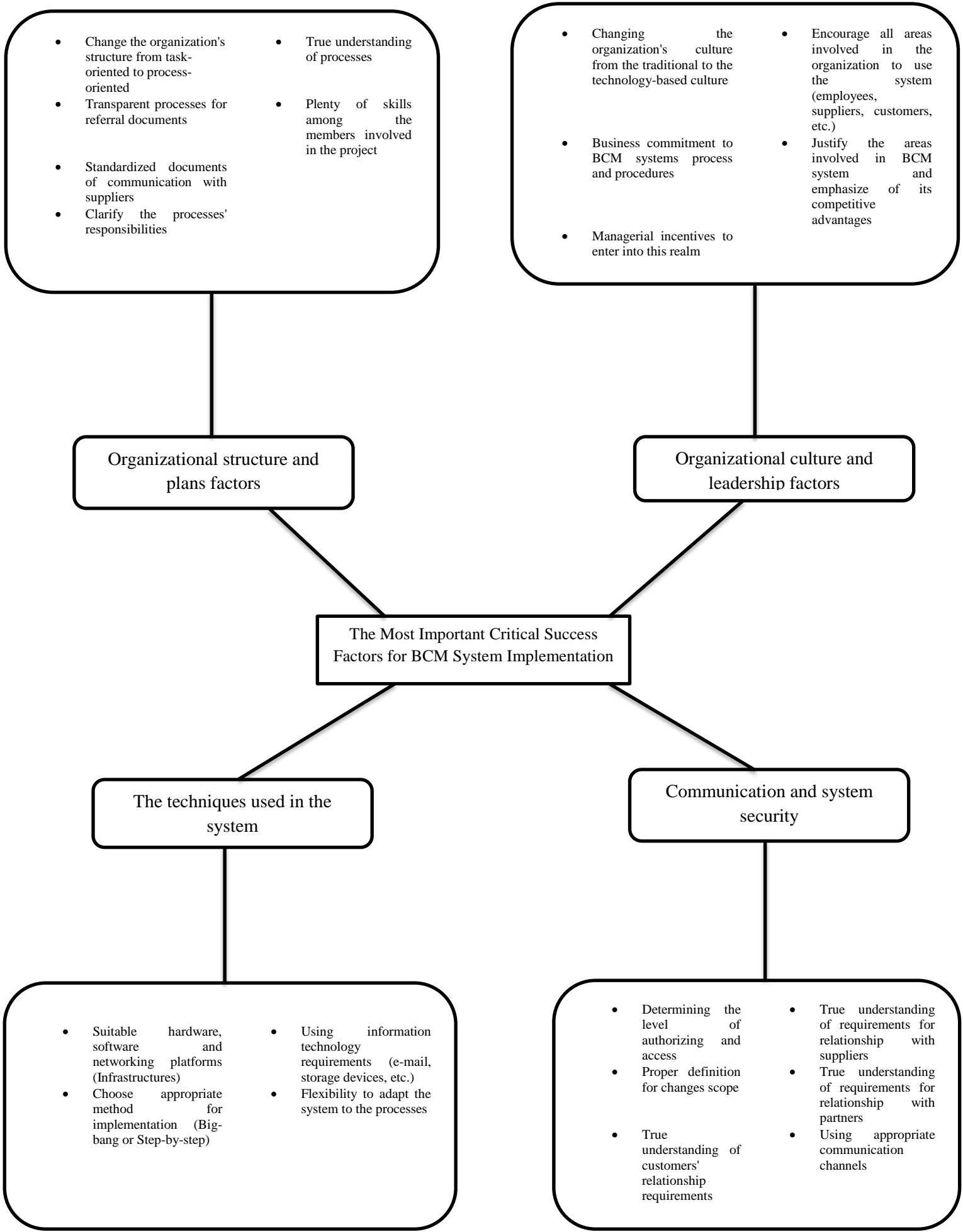

Figure3. The most critical success factors for BCM system implementation

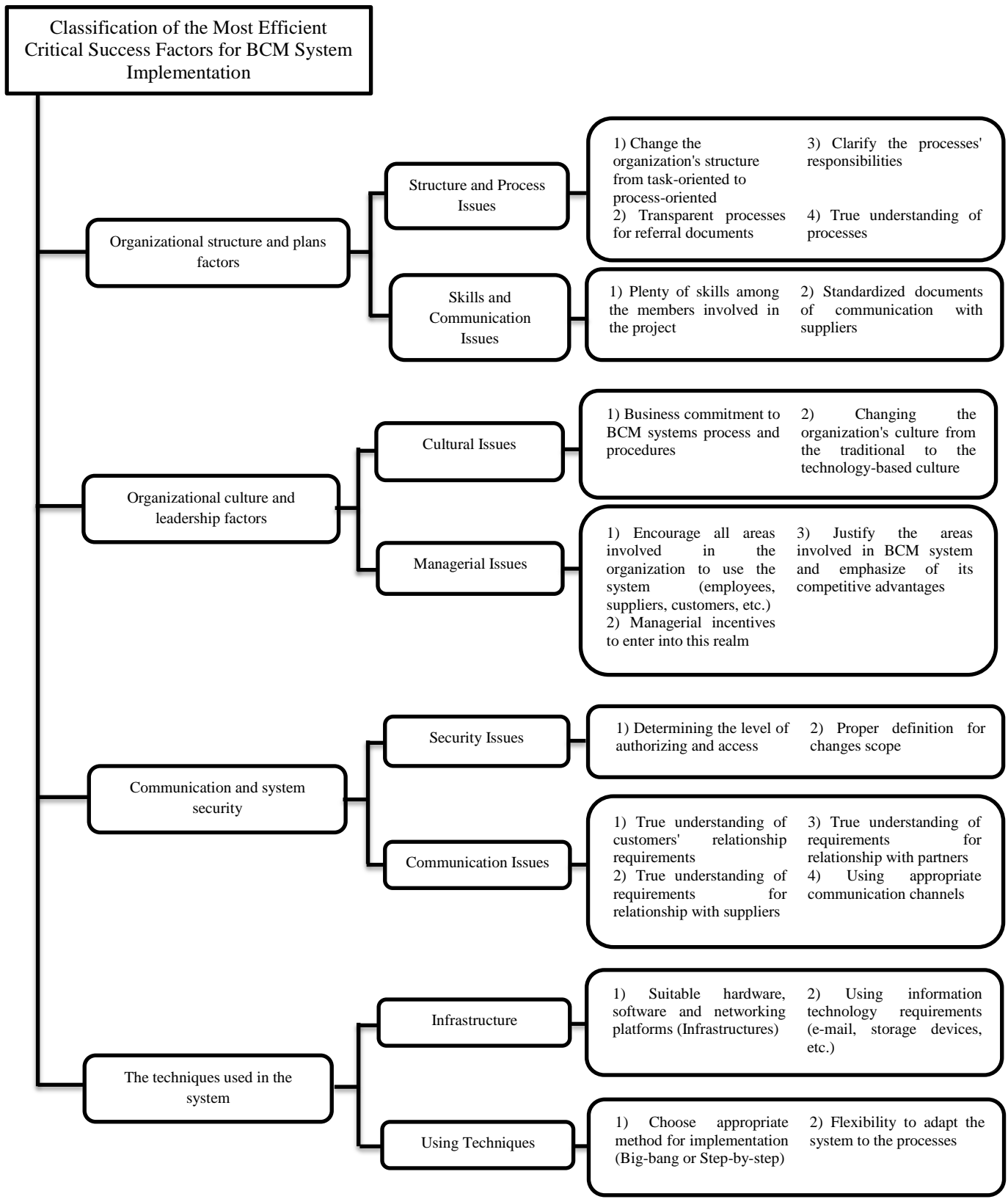

Figure4. 2-Level classification of the most critical success factors for BCM system implementation